\documentclass[reprint]{JASA}

\begin{document}

\title[Measuring pitch extractors]{Measuring pitch extractors' response to frequency-modulated multi-component signals}


\author{Hideki Kawahara}

\email{kawahara@wakayama-u.ac.jp}

\affiliation{Center for Innovative Research and Liaison, Wakayama University, Wakayama, 640-8510 Wakayama, Japan}

\author{Kohei Yatabe}
\affiliation{Department of Electrical Engineering and Computer Science, Tokyo University of Agriculture and Technology, Tokyo, 184-8588 Tokyo, Japan}

\author{Ken-Ichi Sakakibara}
\affiliation{Department of Speech-Language-Hearing Therapy, Health Science University of Hokkaido, Sapporo, 061-0293 Hokkaido, Japan}

\author{Tatsuya Kitamura}
\affiliation{Faculty of Intelligence and Informatics, Konan University, Kobe, 658-8501 Hyogo, Japan}

\author{Hideki Banno}
\affiliation{Department of Information Engineering, Meijo University, Nagoya, 468-8502 Aichi, Japan}

\author{Masanori Morise}
\affiliation{Department of Frontier Media Science, Meiji University, Tokyo, 164-8525 Tokyo, Japan}








\date{\today} 

\begin{abstract}
This article focuses on the research tool for investigating the fundamental frequencies of voiced sounds.
We introduce an objective and informative measurement method of pitch extractors' response to frequency-modulated tones.
The method uses a new test signal for acoustic system analysis.
The test signal enables simultaneous measurement of the extractors' responses.
They are the modulation frequency response and the total distortion, including intermodulation distortions.
We applied this method to various pitch extractors and placed them on several performance maps.
We used the proposed method to fine-tune one of the extractors to make it the best fit tool for scientific research of voice fundamental frequencies.
\end{abstract}


\maketitle


\section{Introduction}
\label{ss:Introduction}
Scientific investigations need reliable and accurate measuring equipment.
Pitch extractors are such measuring equipment for investigating the fundamental frequency of voiced sounds.
However, since the performance comparison report of pitch extractors in 1993 \cite{Titze1993}, no comprehensive report has followed despite spite of the increasing number of new pitch extractors.
We introduce a new performance report of pitch extractors focusing on response to frequency modulations of voiced sounds.
The report uses a method to objectively measure pitch extractors' performance regarding the frequency transfer function, total distortions, and signal-to-noise ratio \cite{kawahara2021icassp,kawahara2021interspeech,kawahara2021APSIPA}.
We measured representative pitch extractors and reported raw data, scientific visualization movies, and characterized pitch extractors on performance maps \cite{kawahara2022IS}.
We propose to use these responses to frequency-modulated multicomponent tones as additional information for evaluating pitch extractors.

This information is also helpful to refine existing pitch extractors.
The measurement and visualization of the responses indicated that one of the extractors shows desirable behavior as a measuring tool for detailed analysis of voice $f_\mathrm{o}$.
We applied the proposed analysis method to fine-tune the extractor (National Institute for Japanese Language and Linguistics (NINJAL) extractor \cite{HidekiKawahara2017,kawahara2017accurate} specially designed to conduct a detailed analysis of the CSJ Japanese spontaneous speech corpus \cite{maekawa03_sspr}) to have the best response bandwidth and gain stability regarding the modulation frequency transfer function of frequency-modulated multi-component signals.

In the following sections, we briefly introduce a specific research question that led to the development of the method proposed in this manuscript.
Then, we outline the objective measurement procedure with theoretical backgrounds.
Followed by descriptions illustrating the principle of operations, we present measurement results of representative pitch extractors and characterize them on performance maps.
Based on the measurement and mapping, we selected one pitch extractor, NINJAL, to fine-tune for detailed analysis of voice $f_\mathrm{o}$.
Finally, we discuss further issues and applications.
We put technical details and information about the tested pitch extractors in appendices and made the tool open-source.

\section{Background and motivation}
\label{ss:Background}
The first author reported the voice-pitch responds to the pitch perturbation of the fed-back voice while the subjects is sustaining a vowel sound in a constant pitch \cite{kawahara1994interactions,kawahara1996icslp}. The experiment used a realtime pitch changer, harmonizer, with MIDI-controlled perturbation pattern made from a pseudo random signal,
named a maximum length sequence (MLS) \cite{schroeder1970synthesis}.
These experiments suggested a consistent compensating response to the perturbation with a relatively short latency (around 100~ms).
Following decades, this led to the altered auditory feedback research focusing on pitch-shift paradigm and response with longer latencies \cite{burnett1997voice}.
This pitch-shift paradigm resulted in many findings \cite{larson2005Jasa,BEHROOZMAND201289,PATEL2016772.e33,larson2016sensory,peng2021causal} and led to their neuronal basis \cite{TOURVILLE20081429,houde2013PNAS,behroozmand2020modulation}.
Unfortunately, possibly because of the technical complexity both in hardware setting and analysis procedures, the response with short latency was less explored \cite{hain2000instructing,zarate2010neural}.

In addition to the issues mentioned above, the procedure introduced in 1994 had fundamental difficulties.
The MLS signal is sensitive to nonlinearity of the tested systems \cite{farina2000simultaneous,stan2002comparison,burrascano2019swept}.
This susceptibility is problematic because biological systems generally consist of non-linear components.
The less sensitive test signal, Swept-sine, is predictable and irrelevant for measuring involuntary response, which is the cause of the short-latency response.
The spectral shaping used in the experiment cannot separate the source of the transients caused by the target system and the shaping artifact.

A new test signal, cascaded all-pass filters with randomized center frequencies and phase polarities (CAPRICEP) \cite{kawahara2021icassp,kawahara2021interspeech,kawahara2021APSIPA}, and side-lobe-lass Gaussian function solved these difficulties. 
The new test signal made it possible to simultaneously measure the linear time-invariant (LTI) response and other responses, including harmonic and intermodulation distortions and random and time-varying responses \cite{kawahara2021icassp}.
Convolution of this new test signal and side-lobe-lass Gaussian function made separation of the system transient possible because the test signal has no transients.
Moreover, we noticed that voice fundamental frequency responds to some frequency modulated tones \cite{larson2005Jasa}.
These phenomena removed the necessity to modulate the fed-back speech sounds in realtime \cite{kawahara2021interspeech,kawahara2021APSIPA}.
This removal of feedback signal processing enables investigations on constituent subsystems, pitch perception, and voice $f_\mathrm{o}$ control, separately and non-invasively.

Under usual test conditions, where subjects can aurally monitor their voice pitch, a combination of a bandpass filter with the target $f_\mathrm{o}$ as its center frequency and instantaneous-frequency analysis is enough to conduct the experiment \cite{kawahara2021interspeech,kawahara2021APSIPA}.
However, under masked auditory feedback conditions, some subjects could not keep the voice fundamental frequency constant.
Sometimes, the voice $f_\mathrm{o}$ deviation exceeded one octave.
This huge deviation made the above-mentioned simple setting fails.
This failure motivated us to test existing pitch extractors regarding as measuring equipment.
We removed the human subject in the voice response measurement setup and let pitch extractors observe the test sounds directly to investigate pitch extractors' performance.

\section{Objective measurement}
\label{ss:ObjMeasurement}
Our target is to observe the fundamental frequency response to auditory stimulation using frequency-modulated speech-like sounds, multi-component harmonic sounds.
We need the pitch extractor to behave like measuring equipment with objective specifications of its functions.
In this manuscript, we introduce a method to measure the frequency modulation transfer function of the pitch extractor.
In addition, we measure spurious responses, including harmonic distortions, intermodulation distortions, and random variations due to noise and other factors.
A new test signal made from CAPRICEP \cite{kawahara2021icassp} is the key for this measurement.

\subsection{CAPRICEP-based simultaneous measurement}
\label{ss:CAPRICEP}
The impulse response of an LTI-system is unique irrespective of test signals.
CAPRICEP-based simultaneous measurement takes advantage of this uniqueness.
High degrees of freedom for designing CAPRICEP signals enabled us to make a mixture of test signals and separate them into orthogonal signals.
Repetitive presentation of the mixture to pitch extractors and the following analysis provide a set of impulse responses to different input signals and observations.
The produced impulse responses are not identical.
The sample mean and the conditional sample variances yield the LTI-response, the non-linear time-invariant distortions, and random and time-varying responses.
In section~\ref{ss:respAnalysis} we introduce details of analysis using CAPRICEP.
Details of the mixed-signal generation and orthogonalization are in Appndix~\ref{ss:capricepDetail}.

\subsection{Measurement procedure}
\label{ss:procedure}
The measurement procedure and supporting tools are implemented using MATLAB R2022a \cite{MATLAB:R2022a} running on a PC (Macbook Pro 14 inch 2021, Apple M1 Max, 64GB).
Tests using deep-learning-based pitch extractors were conducted on different PC (Mac mini 2020, Apple M1, 16GB) using MATLAB R2021a, Updates 5.
We used the live script feature of MATLAB for making visualization movies.

\section{Measurement of pitch extractors}
\label{ss:extractors}
This section describes the underlying signal model, measurement scheme, analysis procedure, and performance indices.
We focus on voiced speech analysis especially sustained vowels.
We model the discrete-time speech signal $s[n]$ as a sum of frequency and amplitude modulated harmonic component and residuals.
\begin{align}
s[n] & = r[n] + \sum_{k = 1}^K a_k[n] \sin\left(\varphi[k] + 2 \pi k \sum_{m=0}^n f_\mathrm{o}[m] t_{\Delta}\right) , \label{eq:signalmodel}
\end{align}
where $f_\mathrm{o}[n]$ represents the instantaneous frequency of the fundamental component.
The symbol $a_k[n]$ represents the instantaneous amplitude of the $k$-th harmonic component.
The symbol $\varphi[k]$ is the initial phase of the $k$-th harmonic component.
The pitch extractors' objective is to estimate the value of $f_\mathrm{o}[n]$ from the observed voiced sound $s[n]$.

\subsection{Measurement scheme}
\label{ss:measuring scheme}
\begin{figure}[ht]
\includegraphics[width=\hsize]{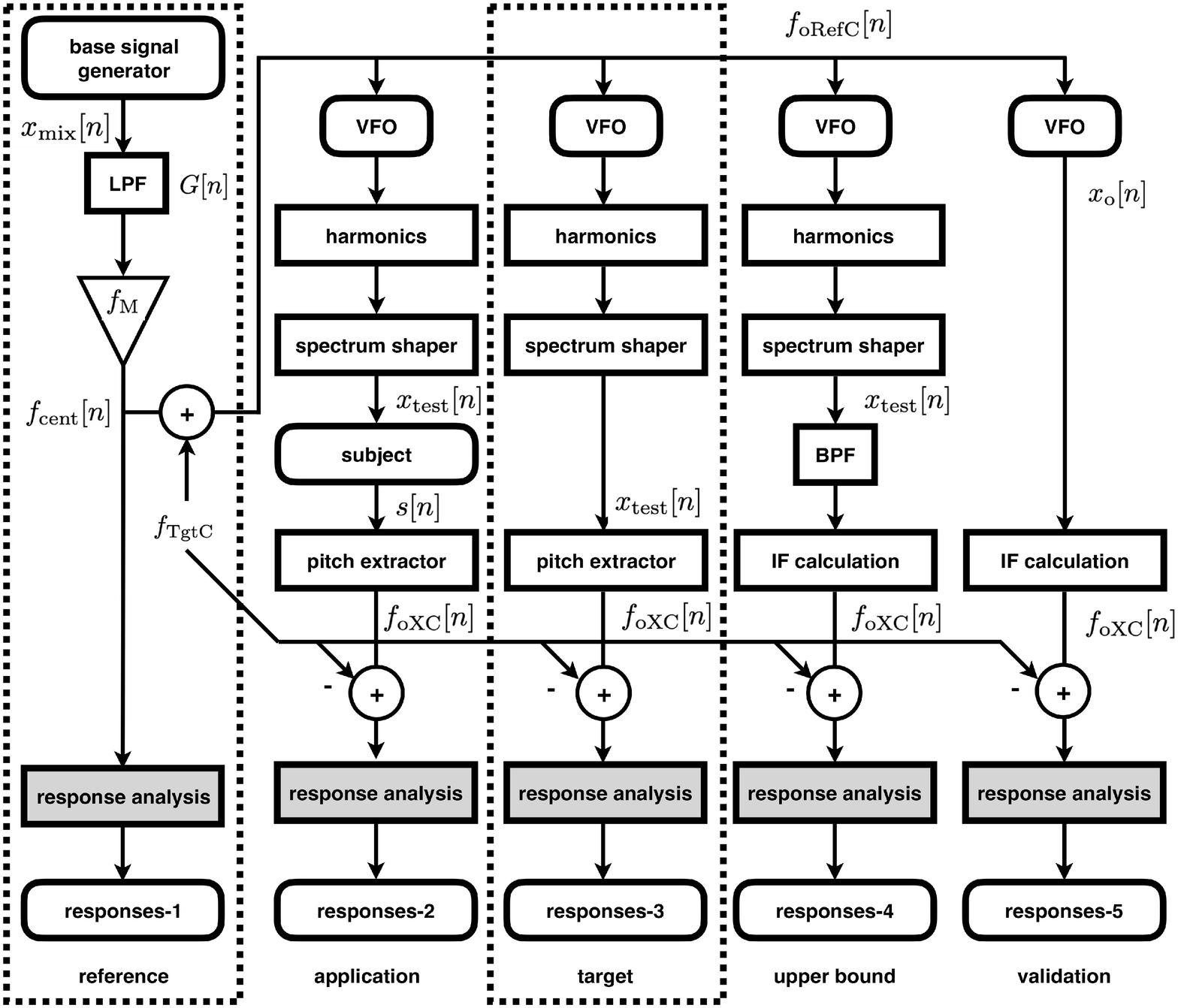}\\
\caption{Measurement scheme and references.}
\label{fig:scheme}
\end{figure}
Figure~\ref{fig:scheme} shows the systems we discuss in this article.
We use the reference bock and the target block to analyze the pitch extractor's responses and performances.
The application block represents the voice response experiment described in the background.
The validation block is to validate the operation of building blocks.
The upper bound block is to check the best possible performance of pitch extractors.

We put symbols of signals discussed in the following section in Fig.~\ref{fig:scheme}.
The following section defines the signals used in this article.

\subsubsection{Test signal definition}
\label{ss:testsignal}

Let $x_\mathrm{mix}[n]$ a base test signal made from three members of CAPRICEP (Appendix~\ref{ss:capricepDetail} provides details.).
We first convolved $x_\mathrm{mix}[n]$ with a side-lobe-less Gaussian $G[n]$ to generate the frequency modulation signal $f_\mathrm{cent}[n]$.
\begin{align}\label{eq:fqmodSignal}
f_\mathrm{cent}[n] & = f_\mathrm{M} \cdot \frac{x_\mathrm{mix}[n] \circ G[n]}{\sigma(x_\mathrm{mix}[n] \circ G[n])} \ \ ,
\end{align}
where the symbol $\circ$ represents convolution and $f_\mathrm{M}$ represents the modulation depth in musical cent.
The function $\sigma(x[n])$ calculates the standard deviation of the signal $x[n]$.

The fundamental frequency of the test signal $f_\mathrm{oRefC}[n]$ represented in cent is defined below.
\begin{align}
f_\mathrm{oRefC}[n] & = f_\mathrm{TgtC} + f_\mathrm{cent}[n] \ \ ,
\end{align}
where $f_\mathrm{TgtC}$ represents the carrier frequency represented in musical cent.
In the following sections, we use $f_\mathrm{o}$ to represent the carrier frequency of the test signal converted from $f_\mathrm{oRefC}[n]$ in musical cent to the linear frequency in Hz.

We use this, $f_\mathrm{o}[n]$, instantaneous frequency of the fundamental component to generate the fundamental component $x_\mathrm{o}[n]$ and harmonic components $x_{\mathrm{h}k}[n]$.
\begin{align}
x_\mathrm{o}[n] & = \sin\left(2 \pi \sum_{m=0}^n f_\mathrm{o}[m] t_{\Delta}\right) \\
x_{\mathrm{h}k}[n] & = \sin\left(2 \pi k \sum_{m=0}^n f_\mathrm{o}[m] t_{\Delta}\right) ,
\end{align}
where we removed initial phase in the signal model of speech from the harmonic components of the test signal\footnote{For psychophysical research test signals with the initial phase setting is useful \cite{patterson1987pulse}. It may also useful in speech coding \cite{kleijn2003signal}.}.
The test signal $x_{\mathrm{test}}[n]$ is a weighted sum of these components.
\begin{align}
x_{\mathrm{test}}[n] & = a_1[n]x_\mathrm{o}[n] + \sum_{k = 2}^K a_k[n] x_{\mathrm{h}k}[n] ,
\end{align}
where we used Japanese vowel spectral shape to define the coefficients $a_k[n], k = 1, \ldots, K$.

\subsubsection{Test signal generation}
\label{ss:tsigGeneration}
Because the target application is a measurement of the fundamental frequency of the sustained voices, we set the length of the test signal to twenty seconds.
We set the sampling frequency to 44100~Hz.
The response to pitch alternation lasts about one second, and we selected the period of the test signal to $2^{16} = 65536$ samples, 1.4861~s.
It allows us to allocate nine segments of the basic unit of the test signal.
The basic unit consists of four possible combinations of three CAPRICEP signals.
We overlap and add nine segments on the time axis to generate the base test signal $x_\mathrm{mix}[n]$ separated by 65536 samples.
Then, the procedure mentioned-above generated the test signal $x_{\mathrm{test}}[n]$.
In the following tests, we set the coefficients constant in time $a_k, k = 1, \ldots, K$.
We made the target frequency, $f_\mathrm{Tgt}$, which is represented in Hz and corresponds to $f_\mathrm{TgtC}$ in musical cent, span from 80~Hz to 400~Hz in 1/48~octave steps.
It resulted in 112~targets.

\subsubsection{Pitch extraction}
\label{ss:pitchExtraction}
We fed the test signal to the pitch extractor of interest.
We prepared an interface function for each pitch extractor as a MATLAB function.
Appendix~\ref{ss:InterfaceImplementation} shows details.
We can test any pitch extractor of interest by writing a similar interface function.

\subsubsection{Response analysis}
\label{ss:respAnalysis}
We converted the extracted fundamental frequency $f_\mathrm{oX}[m]$, where $m$ represents the analysis frame index, to a discrete-time signal $f_\mathrm{oXC}[n]$ of 44100~Hz sampling using linear interpolation represented in the musical cent.
Then, we subtracted the target frequency $f_\mathrm{TgtC}$ to make the input to response analysis.

The CAPRICEP-based analysis procedure generates six responses $u_\mathrm{L}^{(p)}[n]$ with the length 65536, where $p$ represents the segment identifier.
They are calculated from six periodic segments aligned on the time axis without overlaps.
The analysis procedure also generates three set of four responses $u_\mathrm{S}^{(p,j,k)}[n]$ for each segment, where $j, j=1,2,3$ represents the identifier of the orthogonal sequence and $k, k = 1,2,3,4$ represents the identifier of the combination.
The length of $h_\mathrm{S}^{(p,j,k)}[n]$ is 16384 samples.

We calculated discrete Fourier transform of $u_\mathrm{L}^{(p)}[n]$ and $u_\mathrm{S}^{(p,j,k)}[n]$.
We represent them as $U_\mathrm{L}^{(p)}[k]$ and $U_\mathrm{S}^{(p,j,k)}[k]$.
We also calculate discrete Fourier transform of the reference signal $u_\mathrm{RefL}[n]$ and $u_\mathrm{RefS}[n]$.
We represent them as $U_\mathrm{RefL}[k]$ and $U_\mathrm{RefS}[k]$.

The following equation defines the modulation-frequency transfer function $H[k]$ of the tested pitch extractor.
\begin{align}
H[k] & = \frac{1}{6} \sum_{p=1}^{6} H^{(p)}[k]  \\
 \mbox{where} &  \ \ \ H^{(p)}[k] = \frac{U_\mathrm{L}^{(p)}[k]}{U_\mathrm{RefL}[k]} . \nonumber
\end{align}

The following equation defines the sample variance of random and time-varying response $\sigma_\mathrm{TV}^2[k]$.
\begin{align}
\sigma_\mathrm{TV}^2[k] & = \frac{1}{5} \left|\sum_{p=1}^{6} H^{(p)}[k] - H[k] \right|^2 ,
\end{align}
where the denominator $5=6-1$ is for the adjustment of the degrees of freedom.

We define the variation of the transfer function for each CAPRICEP signal combination $\sigma_\mathrm{nLTI}^{(p,j)}[k]$ using the following equation.
\begin{align}
\sigma_\mathrm{nLTI}^{(p,j)}[k] & = \frac{1}{2} \sum_{i = 1}^3 \left|H^{(p,i,j)}[k] - H^{(p,j)}[k]  \right|^2 \\
\mbox{where} &  \ \ \ H^{(p,j)}[k] = \frac{1}{2} \sum_{i = 1}^3 H^{(p,i,j)}[k]  \nonumber \\
  &  \ \ \ H^{(p,i,j)}[k] = \frac{U_\mathrm{S}^{(p,i,j)}[k]}{U_\mathrm{RefS}[k]} . \nonumber
\end{align}

Then, we define the sample variance of the non-linear time-invariant response $\sigma_\mathrm{nLTI}^2[k]$ using the following equation.
\begin{align}
\sigma_\mathrm{nLTI}^2[k] & = \frac{1}{6 \times 4}\sum_{p-1}^6 \sum_{j=1}^4 \left(\sigma_\mathrm{nLTI}^{(p,j)}[k]\right)^2 .
\end{align}

\subsection{Performance indices}
\label{ss:performance}
We introduce four performance indices for characterizing pitch extractors.
They are bandwidth $B_\mathrm{w}$, signal to noise ratio $SNR_\mathrm{FM}$, standard deviations of gain change in frequency $SD_\mathrm{fd}$ and in time $SD_\mathrm{td}$.
The following sections illustrate these indices with equations and analysis results.

\subsubsection{Bandwidth}
We use the second-order moment to define the bandwidth index $B_\mathrm{w}$.
First, we define the LTI power function $P_\mathrm{LTI}[k]$ as the squared absolute value of the transfer function $P_\mathrm{LTI}[k] = |H[k]|^2$, and define the total disturbance power function $P_\mathrm{TD}[k]$ as the sum of the non-linear time-invariant response, and random and time-varying response.
We select a set of discrete frequency index $k$ to define the evaluation set $\Omega = \{k | k <= k_\mathrm{B} \}$ where $k_\mathrm{B}$ is the lowest discrete frequency satisfies $P_\mathrm{LTI}[k] < P_\mathrm{TD}[k]$.
Then, following equation defines $B_\mathrm{w}$.
\begin{align}
B_\mathrm{w} & = \sqrt{\frac{\sum_{k \in \Omega} (f[k])^2 P_\mathrm{LTI}[k]}{\sum_{k \in \Omega} P_\mathrm{LTI}[k]} } ,
\end{align}
where $f[k]$ represents the function that maps the discrete frequency to the frequency (unit: Hz).

\subsubsection{Signal to noise ratio}
We define the Signal to noise ratio SNR using this bandwidth.
The following equation defines $SNR_\mathrm{FM}$.
\begin{align}
SNR_\mathrm{FM} & = 10 \log_{10} \left( \frac{\sum_{k \in \Omega_{B_\mathrm{w}}} P_\mathrm{LTI}[k]}{\sum_{k \in \Omega_{B_\mathrm{w}}} P_\mathrm{TD}[k]}\right) ,
\end{align}
where $\Omega_{B_\mathrm{w}}$ represents the set of discrete frequencies defined by $\Omega_{B_\mathrm{w}} = \{ k |f[k] < B_\mathrm{w}  \}$.
Note that frequency 0 is not a member of $\Omega_{B_\mathrm{w}}$.

\subsubsection{Gain change}
The gain representation of the modulation transfer function is a function of the modulation frequency and the signal's fundamental frequency.
For measuring equipment, the gain should be constant for pre-defined frequency ranges.
We define two performance indices for modulation frequency and fundamental frequency.

The following equation defines the standard deviation of gain change in modulation frequency $SD_\mathrm{fd}$.
\begin{align}
SD_\mathrm{fd} & = \sqrt{\frac{1}{N_f \Delta f_\mathrm{x}}\sum_{k \in \Omega_{10}}\left| 10 \log_{10}\left(
   \frac{P_\mathrm{LTI}[k+1]}{P_\mathrm{LTI}[k]} \right)\right|^2} ,
\end{align}
where $\Omega_{10}$ represents the set of discrete modulation frequencies $\Omega_{10} = \{k| 0< f[k], f[k+1] < 10 \}$ (unit: Hz).
The symbol $N_f$ represents the cardinal number of $\Omega_{10}$ and $\Delta f_\mathrm{x}$ represents the frequency difference of neighboring discrete frequencies.
The unit of this index is dB/Hz.

Let $\overline{P_\mathrm{LTI}}(f_\mathrm{o}[n])$ represents the average gain in $\Omega_{10}$ for a test signal generated using the $n$-th fundamental frequency $f_\mathrm{o}[n]$ Hz.
Then, following equation defines the standard deviation of gain change in modulation frequency $SD_\mathrm{fd}$.
\begin{align}
\!\!SD_\mathrm{t_d} & = \sqrt{\frac{1}{N_t \Delta f_\mathrm{o}}\!\!\sum_{n \in \Omega_{N}}\!\!\left| 10 \log_{10}\!\!\left(
   \frac{\overline{P_\mathrm{LTI}}(f_\mathrm{o}[n+1])}{\overline{P_\mathrm{LTI}}(f_\mathrm{o}[n])}\! \right)\right|^2} ,
\end{align}
where $\Omega_{N}$ represents the set of target frequencies $\Omega_{N} = \{k| f_L \le f_\mathrm{o}[n], f_\mathrm{o}[n] \le f_H \}$ (unit: semitone).
The symbol $f_L$ represents the lower bound, and $f_H$ represents the upper bound of the target frequencies.
The symbol $N_t$ represents the cardinal number of $\Omega_{N}$ and $\Delta f_\mathrm{o}$ represents the step size of the target frequencies (unit: semitone).
The unit of this index is dB/semitone.

\section{Results}
We analyzed twenty pitch extractors.
Refer to Appendix~\ref{ss:testedExtractor} for the descriptions of the tested pitch extractors.
This section shows representative excerpted responses and summary maps of the statistical analysis results.

\subsection{Frequency response}
\label{ss:freqResp}
\begin{figure}[tbp]
\includegraphics[width=\hsize]{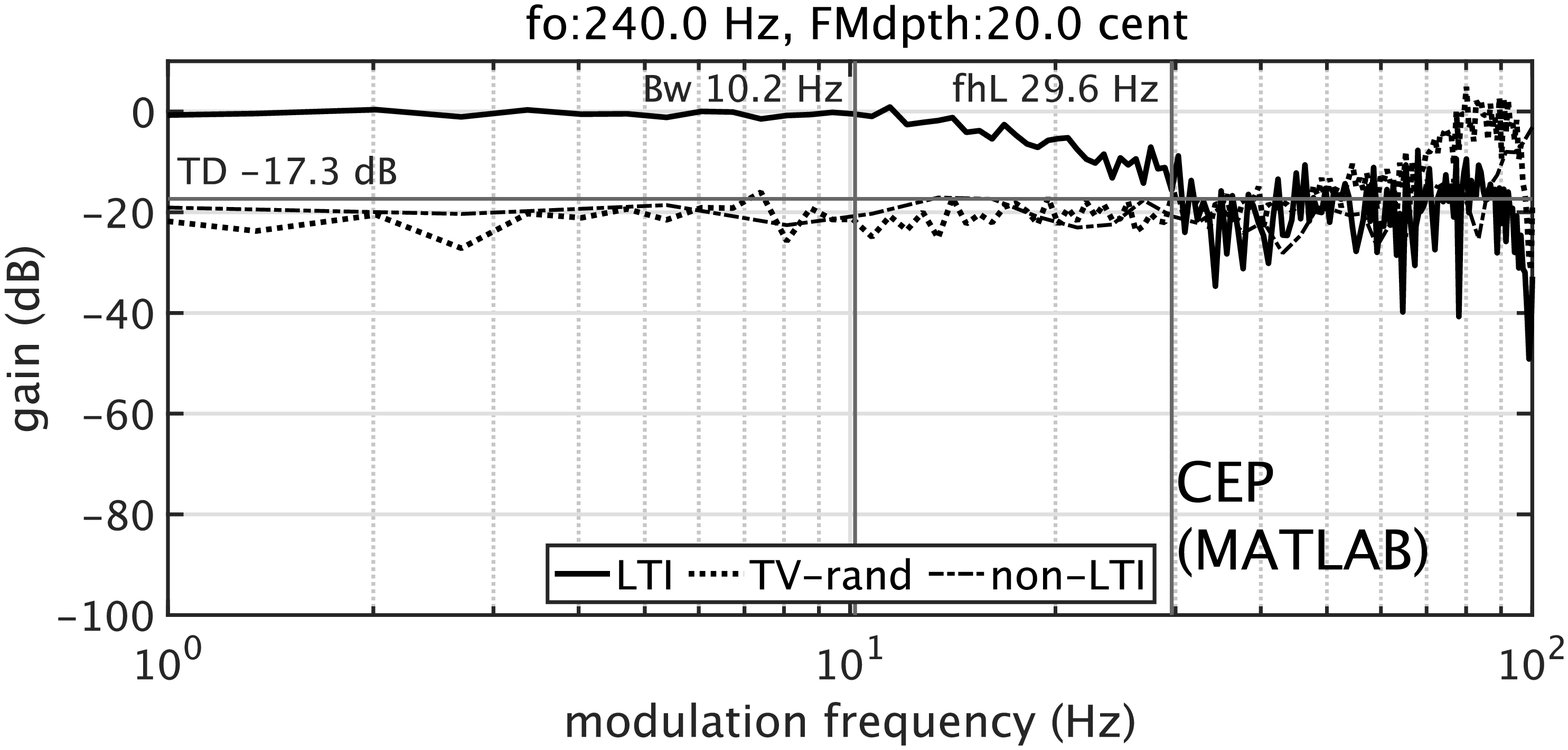}
\caption{Response of the cepstrum-based pitch extractor.\label{fig:cepResponse}}
\end{figure}

\begin{figure}[tbp]
\includegraphics[width=\hsize]{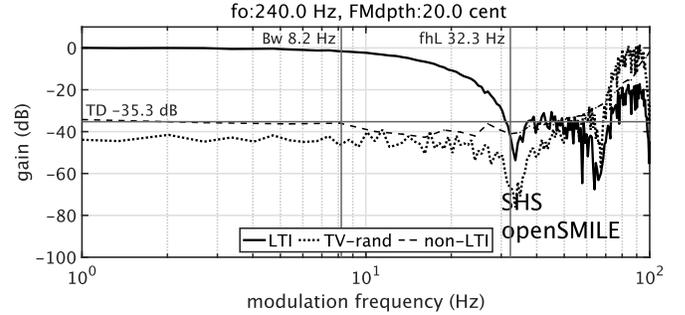}
\caption{Response of the subharmonic sampling method in openSMILE.\label{fig:oSMILEResponse}}
\end{figure}

\begin{figure}[tbp]
\includegraphics[width=\hsize]{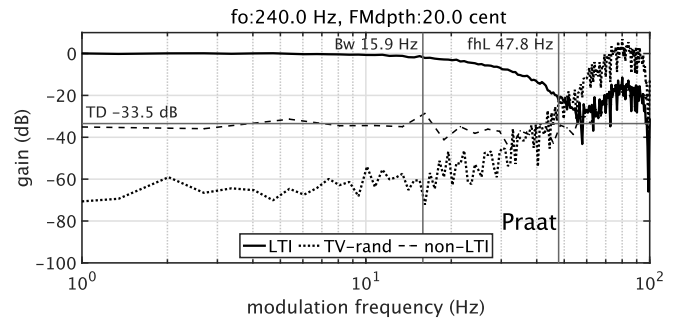}
\caption{Response of the Praat's default pitch extractor.\label{fig:praatResponse}}
\end{figure}

\begin{figure}[tbp]
\includegraphics[width=\hsize]{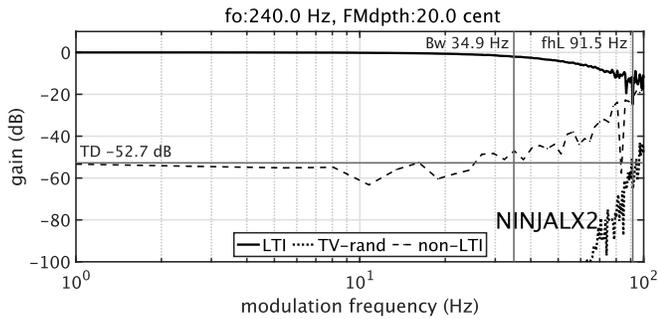}
\caption{Response of the NINJAL pitch extractor after fine-tune.\label{fig:ninjalResponse}}
\end{figure}

Figures~\ref{fig:cepResponse},\ref{fig:oSMILEResponse},\ref{fig:praatResponse},\ref{fig:ninjalResponse} show analysis results of several pitch extractors.
The thick solid line with legend \textsf{LTI} represents the LTI response $P_\mathrm{LTI}[k]$.
The thick dotted line with legend \textsf{TV-rand} represents the sample variance of the random and time-varying response $\sigma_\mathrm{TV}^2[k]$.
The dash-dot line with legend \textsf{non-LTI} represents the smaple variance of the non-linear time-invariant response $\sigma_\mathrm{nLTI}^2[k]$.

The vertical line annotated by \textsf{fhL} represents the boundary frequency that defines $\Omega$.
The vertical line annotated by \textsf{Bw} represents the bandwidth $B_\mathrm{w}$.
The horizontal line annotated by \textsf{TD} represents the sum of the non-linear time-invariant response and the random and time-varying response.
In the title of the figure, \textsf{fo} represents the target carrier frequency $f_\mathrm{o}$ and \textsf{FMdpth} represents the modulation depth $f_\mathrm{M}$.

Figure~\ref{fig:cepResponse} shows the result of a cepstrum-based pitch extractor \cite{noll1967jasa}.
The total distortion is the highest among four plots.
The source of the distortion is the quantization of the estimated fundamental frequency.
This method find the peak position on the discrete quefrency bins separated by the signal sampling interval.
We found that some of the existing pitch extractors' implementation suffers from this quantization.

There are many ways to circumvent this quantization effect.
Figure~\ref{fig:oSMILEResponse} is such an example.
The pitch extractor is one configuration of a popular software tool, openSMILE \cite{eyben2010opensmile,eyben2010opensmile2}, developed for paralinguistic research \cite{schuller2013computational}.
The pitch extractor uses a classical subharmonic summation-based method \cite{hermes1988measurement} while does not suffer from the quantization effect mentioned above.

Figure~\ref{fig:praatResponse} shows the result of the default pitch extractor of a popular software tool for doing linguistics using computers, Praat \cite{boersma2011praat}.
The performance of this pitch extractor is the best of the tested pitch extractors other than ours.
The documentation \cite{boersma2011praat} of the tool and personal communication with an author of Praat suggested careful implementation of the underlying algorithm \cite{boersma1993accurate}.

Figure~\ref{fig:ninjalResponse} shows the result of an instantaneous-frequency-based pitch extractor NINJAL \cite{HidekiKawahara2017,kawahara2017accurate}.
The total distortion is the lowest, and the bandwidth is the highest.
They are close to the upper bound block in Fig.\ref{fig:scheme}.
The method has an automatic selection mechanism for the best bandpass filter.
The analysis rate also contributes to significantly reducing the random response level.
NINJAL calculates the fundamental frequency at the audio sampling rate\footnote{%
In this test, automatic downsampling of NINJAL kicked in. The analysis rate was 44100/6~Hz.}
Note that we fine-tune the smoothing time constant of the post-processing in NINJAL from 40~ms to 10~ms to get this result.


The measurement of twenty pitch extractors produced 2240 plots.
We sequenced each pitch extractor's plots according to the fundamental frequency and converted the aligned plots to a movie to quickly grasp the extractor's behavior.
We also gathered sixteen movies and made a movie to inspect them at once.
These movies helped us to make the performance indices described above\footnote{See Supplementary material at [URL will be inserted by AIP] for [the gathered sixteen movies].}.

\subsection{Performance map}
\label{ss:performanceMap}
This section locates tested pitch extractors on maps using derived performance indices.
We made two maps.
The first map uses the bandwidth and SNR.
The second map uses gain variations on the modulation frequency and fundamental frequency axes.

\begin{figure}[ht]
\begin{center}
\hfill
\includegraphics[width=0.9\hsize]{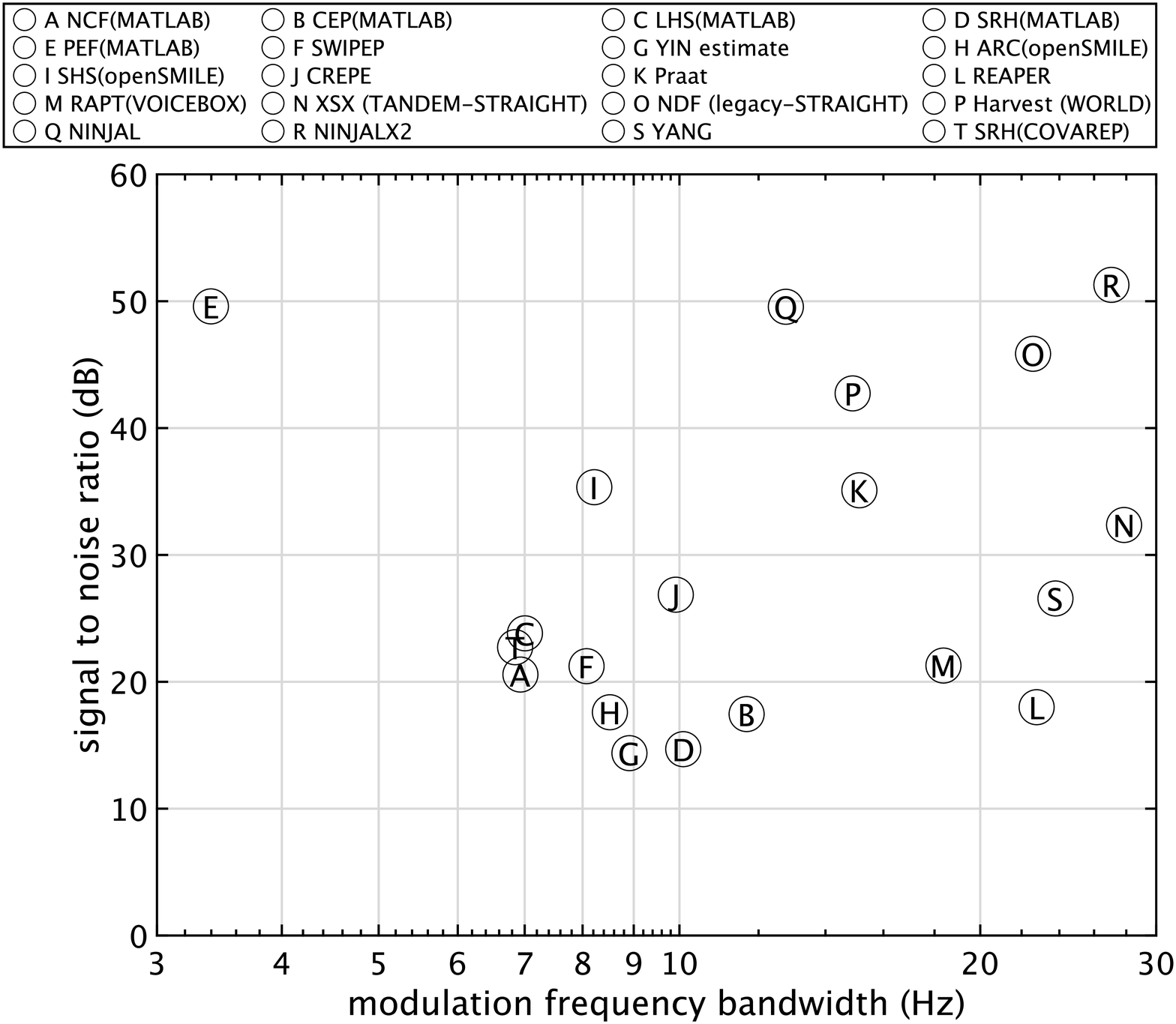}\\
\caption{Pitch extractors location on the bandwidth-SNR map.\label{bwsdmap}}
\end{center}
\end{figure}
Figure~\ref{bwsdmap} is a scatter plot of the pitch extractors on the modulation frequency bandwidth and SNR plane.
Each circle with a character inside represents each pitch extractor.
The plot's legend provides a list of extractors with the symbol characters.
Note that pitch extractors \textsf{N} to \textsf{S} are ours (One of the authors coded them.).
The design target for voice pitch measuring equipment is wider bandwidth and a higher signal-to-noise ratio in the upper right corner of Fig.~\ref{bwsdmap}.

\begin{figure}[ht]
\begin{center}
\hfill
\includegraphics[width=0.9\hsize]{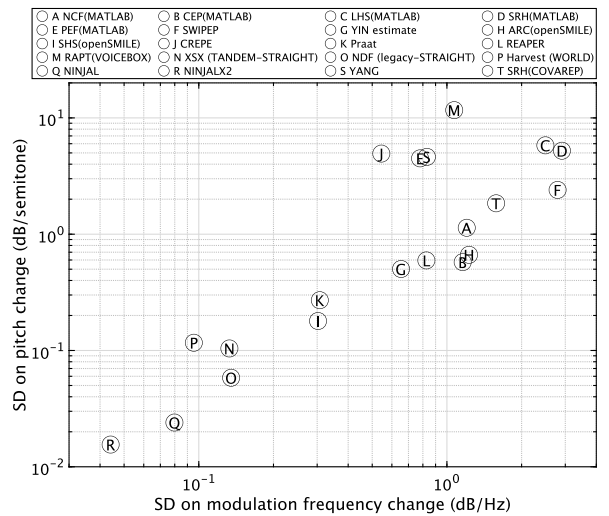}\\
\caption{Pitch extractors location on the gain changes on the modulation frequency-pitch map.\label{gainchangemap}}
\end{center}
\end{figure}
Figure~\ref{gainchangemap} is a scatter plot of the pitch extractors regarding gain variation on the modulation frequency dependence to the fundamental frequency dependence plane.
The symbols and legend are the same as Fig.\ref{bwsdmap}.
For voice pitch measuring equipment, smaller variation in both aspects is desirable.
That is the lower-left corner of Fig.~\ref{gainchangemap}.

Inspection of Figs.~\ref{bwsdmap} and \ref{gainchangemap} indicates that pitch extractors \textsf{N} to \textsf{R} form a high-performance cluster.
The extractor \textsf{R} is the result of fine-tuning of \textsf{Q}.
We changed the smoothing time constant (40~ms in \textsf{Q}) to 10~ms to expand the bandwidth.
This change slightly improved SNR and reduced the gain variations.
In other words, this change turned the NINJAL pitch extractor into the best fit tool for investigating the fundamental frequency of sustained voiced sounds.


\section{Discussion}
\label{ss:Discussion}

The proposed performance indices are supplemental to existing performance indices.
Existing pitch extractors are designed and tuned to their target applications, for example speech recognition, and speech synthesis under different conditions.
The modulation frequency transfer function and other responses used in this manuscript are relevant for investigating sustained voiced sounds recorded in quiet laboratory conditions.
Relatively high-performance of our extractors may depends on this difference of target applications.

The underlying signal model causes performance differences.
The assumed signal model for this measurement is a variant of the sinusoidal model \cite{mcaulay1986speech}.
We need to explore pitch extractors assuming the other signal models, especially excitation-based models \cite{kadiri2021extraction}.
We also need to explore relations to deep-learning-based pitch extractors.
Their underlying models are implicit and tuned through the learning process to improve their target applications' performance.
These differences probably contribute to the relatively poor performance of \textsf{CREPE}, which is a deep learning-based pitch extractor.

The proposed tool needs further investigation.
We only used spectral shaping with the Japanese vowel /a/.
It needs to test other vowels, including foreign ones.
The modulation power spectrum of the modulation signal $f_\mathrm{cent}[n]$ is different from that of natural voices \cite{Titze1993}.
We need to test pitch extractors when nonlinearity is not negligible, using test signals having the same acoustic parameters as the voiced sounds in question.

Even with these issues, the proposed measuring tool and the performance indices introduce a new point of view on the old and still hot topic, pitch extraction.
We open-sourced the tool and provided an easy means to apply the tool to any pitch extractors of interest.
A brief description is in Appendix~\ref{ss:InterfaceImplementation}.

\section{Conclusion}
\label{ss:Conclusion}
We introduced an objective measurement method of pitch extractors' response to frequency-modulated multi-component tones.
We introduced performance indices based on the method.
The proposed performance indices provide new and supplemental means for existing performance measures.
We measured representative pitch extractors and selected one of them for further refinement to make it a valuable tool for investing voice $f_\mathrm{o}$ response to auditory stimulation.
We open-sourced the proposed tool to make researchers easily select and tune pitch extractors for their research purposes.

\begin{acknowledgments}
We appreciate Heiga Zen of Google Brain and Kikuo Maekawa of NINJAL for inviting the first author for their projects. Participating in their projects enabled us to develop the reference pitch extractor NINJAL. The naming is an acknowledgment for the Center for Corpus Development, NINJAL.
This research was supported by Grants in Aid for Scientific Research (Kakenhi) by JSPS numbers, JP18K00147, JP18K10708, JP19K21618, JP21H03468, JP21H04900, and JP21H00497.
\end{acknowledgments}

\appendix
\section{CAPRICEP-based measurement}
\label{ss:capricepDetail}
\begin{figure}[tbp]
    \centering
    \includegraphics[width=\hsize]{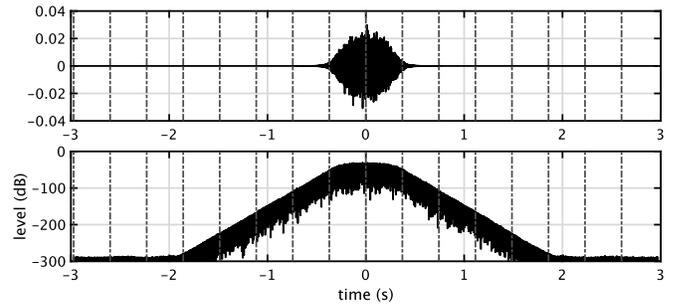}
    \caption{Waveform and level of an unit-CAPRICEP.
    Vertical dash-dot lines represents allocation interval.\label{fig:unitCapricep}}
\end{figure}

An element of CAPRICEP signals is an all-pass filter.
We call it a unit-CAPRICEP.
The transfer function of unit-CAPRICEP is a product of second-order IIR filters.
Using two random numbers, we allocate each all-pass filter's transfer function and its complex conjugate.
The allocation rule is a generalized version of the rule of velvet noise\cite{valimaki2013ieetr}.
The generalization enabled the flexible design of the envelope shape of unit-CAPRICEP\cite{kawahara2021icassp}.
Figure~\ref{fig:unitCapricep} shows an example shape and level of unit-CAPRICEP.

We select three elements $x_\mathrm{CP}^{(1)}[n], x_\mathrm{CP}^{(2)}[n], \mbox{and} x_\mathrm{CP}^{(3)}[n]$ to prepare three sequences.
Then, we generate three base sequences $x_\mathrm{B}^{(1)}[n], x_\mathrm{B}^{(2)}[n], \mbox{and} x_\mathrm{B}^{(3)}[n]$ using following equation.
\begin{align}
   x_\mathrm{B}^{(k)}[n] & = \sum_{m \in \mathbb{Z}} B[m\bmod 4, k] \ 
   x_\mathrm{CP}^{(k)}[n + m N_p] \\ 
 B & = \left[
   \begin{array}{rrr}
      1  & 1 & 1 \\
      1 & -1 & 1 \\
      1 & 1 & -1\\
      1 & -1 & -1
   \end{array}
   \right] ,
\end{align}
where $N_p$ represents the allocation interval in samples.
The sum of these three sequences $x_\mathrm{B}^{(k)}[n], (k = 1, 2, 3)$ is the base test signal $x_\mathrm{mix}[n]$ in Eq.(\ref{eq:fqmodSignal}).

\begin{figure}[tbp]
    \centering
    \includegraphics[width=\hsize]{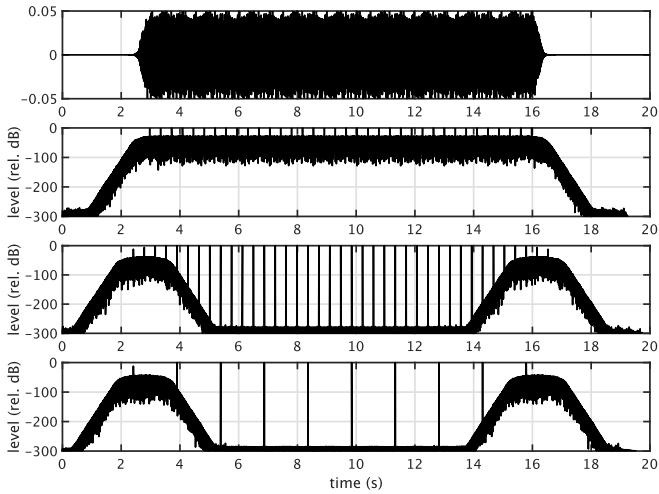}
    \caption{Mixed signal, recovered signal, orthogonalized signal, and extended signal.\label{fig:orth}}
\end{figure}
Figure~\ref{fig:orth} shows signal processing process starting from the mixed signal, the base test signal $x_\mathrm{mix}[n]$ in Eq.(\ref{eq:fqmodSignal}).
Convolution of time reversed version of each element $x_\mathrm{{CP}\ast}^{(k)}[n] = x_\mathrm{CP}^{(k)}[-n]$ and the base test signal recoveres periodically allocated pulses with the pololity defined by $B$.
The resulted signal also consists of temporally spread noise-like cross-correlation between the other sequences.
That is the second plot of Fig.~\ref{fig:orth}.

Periodic shift and add operation using $B$ as coefficient completely cancel these correlations and recover three orthogonal sequences.
This procedure produces four segments consisting of a unit pulse for each sequence.
In other words, the procedure provides twelve impulse responses ($3 \times 4$) calculated using different signals.
The third plot of Fig.~\ref{fig:orth} shows one of these orthogonalized signals.
Note that the cross-correlation level is around -300~dB.

Finally, adding three orthogonalized sequences with 1/4, 1/4, and 1/2 weights provides a segment of length $4 N_p$ consisting of a unit impulse.
The bottom plot of Fig.~\ref{fig:orth} shows this extended signal.
Therefore, the CAPRICEP-based measuring procedure simultaneously measures thirteen impulse responses for one period.
The base test signal in Fig.~\ref{fig:orth} has six periods with enough suppression of crosscorrelation.
Therefore, this base test signal provides 72 ($12 \times 6$) impulse responses and six extended impulse responses.

We take advantage of the base signal periodicity (the period is $4N_p$) and derive an efficient algorithm based on discrete Fourier transform\cite{kawahara2022IS}.
The proposed tool implements this algorithm using the built-in fast Fourier transform (FFT).
The FFT buffer size does not need to be an integer power of two for modern computers because of advancements in algorithms and special-puropse instruction sets, and vector processing mechanisms.

\section{Tested pitch extractors}
\label{ss:testedExtractor}
We found pitch extractors based on the same algorithm showed different performance depending on their detailed implementation.
We listed tested pirch extractors with information rearding their implementation, version, and source locations.

\subsection{MATLAB function (LHS,CEP,SRH,LCF,PEF,CREPE) }
Scientific computing environment MATLAB has builtin functions for pitch extraction.
They consists of classical methods (\textsf{CEP}: cepstrum-based method \cite{noll1967jasa}, \textsf{LCF}: LPC-based method \cite{atal1972automatic}, and \textsf{LHS}: harmonic summation-based method \cite{hermes1988measurement}), and recent methods (\textsf{PEF}: \cite{Gonzalez2014PEFAC} and \textsf{SRH}: summation of residual harmonics \cite{drugman11interspeech}).
In addition to these, \textsf{Deep Leaning Toolbox} has a deep learning-based method \textsf{CREPE} \cite{kim2018crepe}.
\subsection{YIN (ECKF)}
\textsf{YIN} \cite{de2002yin} is an absolute difference-based method originally implemented using MATLAB and C.
The original implementation is not compatible with recent MATLAB versions.
Therefore, we used a variant implementation in the extended complex Kalman filter-based pitch tracker \textsf{ECKF} \cite{das2020improved}.
\subsection{SWIPEP }
\textsf{SWIPEP} \cite{camacho2008jasa} uses a harmonic model-based approach originally implemented in MATLAB.
It estimates the fundamental frequency of the best matching sawtooth signal.
We used the extended version SWIPE' listed in the thesis \cite{camacho2007swipe}.
\subsection{RAPT (VOICEBOX)}
\textsf{RAPT} \cite{talkin1995robust} for robust processing uses a multi-stage autocorrelation-based method followed by post processing.
We used implementation in \textsf{VOICEBOX}: Speech Processing Toolbox for MATLAB \cite{brooksVoicevox}.
\subsection{SRH (COVAREP)}
\textsf{SRH} \cite{hermes1988measurement} is a harmonic summation-based method.
\textsf{COVAREP} is a repository of speech processing tools \cite{Degottex2014icassp}.
We used MATLAB implementation of \textsf{SRH} in COVAREP repository \cite{covarep}.
\subsection{REAPER }
\textsf{REAPER} simultaneously estimate GCI (Glottal Closure Instant), V/UV (voiced or unvoiced), and pitch.
We used the open-source implementation \cite{googleREAPERgit}.
\subsection{Praat }
\textsf{Praat} is a popular tool for doing phonetics using computers \cite{boersma2011praat}.
We used recommended procedure ``Sound: To Pitch...'' which uses the autocorrelation-based method \cite{boersma1993accurate} with the default setting.
We used the latest version v.6.2.10 \cite{boersma2011praat}.
\subsection{openSMILE (SHS, ACF)}
\textsf{openSMILE} is widely applied in automatic emotion recognition for affective computing \cite{eyben2010opensmile}. 
It has two configuration files for pitch analysis, \textsf{prosodyShs.conf} which uses the sub-harmonic-sampling method \textsf{SHS}, and \textsf{prosodyAcf.conf} which uses an autocorrelation and cepstrum-based method \textsf{ACF}.
We used openSMILE version 3.0.1 for macOS \cite{openSmile3}.
\subsection{STRAIGHT (NDF, XSX) }
\textsf{STRAIGHT} consists of two VOCODER packages, legacy-STRAIGHT \cite{kawahara1999spcom} and TANDEM-STRAIGHT \cite{kawahara2008icassp}.
They use $f_\mathrm{o}$ adaptive spectral envelope recovery.
Their $f_\mathrm{o}$ adaptive procedure led to development and refinement of dedicated pitch extractors, \textsf{NDF} \cite{kawahara05_interspeech} for legacy-STRAIGHT, and \textsf{XSX} \cite{kawahara08_spkd} for TANDEM-STRAIGHT.
They are implemented using MATLAB.
The legacy-STRAIGHT and NDF is open-source \cite{legacySTRAIGHT} since 2017.
\subsection{Harvest (WORLD)}
A high-quality VOCODER \textsf{WORLD} \cite{morise2016world} also use $f_\mathrm{o}$ adaptive spectral envelope recovery.
The latest pitch extractor for WORLD is Harvest \cite{morise17b_interspeech}.
We used MATLAB implementation version 0.2.4.
\subsection{NINJAL, NINJALX2 and YANG }
This pitch extractor \textsf{NINJAL} \cite{HidekiKawahara2017,kawahara2017accurate} is a refined version of its predecessor \textsf{YANG} \cite{Kawahara2016}.
They use a log-linear filter-bank and their instantaneous frequency and residual levels of outputs.
We set the smoothing length parameter to 10~ms (named \textsf{NINJALX2}) and 40~ms (named \textsf{NINJAL}: default).
The setting of \textsf{NINJALX2} is the result of fine-tuning enabled by the proposed objective measurement.
They are implemented in MATLAB.
They are open-source and accessible on \textsf{YANG} \cite{yangVocoder} and \textsf{NINJAL} \cite{ninjalGit}.


\section{Interface to pitch extractors}
\label{ss:InterfaceImplementation}

\begin{figure}
{\scriptsize
\begin{verbatim}
function output = pitchCEP(xa, fs)
% Interface function for pitch extractor of Noll's CEPSTRUM
%   output = pitchCEP(xa, fs)
%     Use the function name "@pitchCEP" for the argument 
%     of the evaluator
%
% Augment
%   xa   : test signal with CAPRICEP FM and simulates
%          the spectrum of Japanese vowel /a/ 
%   fs   : sampling frequency (Hz)
% Output
%   output : structure varialbe with the following fields
%      fo   : extracted fundamental frequency (Hz)
%      tt   : discrete temporal locations of fo measurement (s)
%      titleStr    : string for the first item of the figure title
%      filePrefix  : string for the beggining of the output files

% LICENSE: refer to LICENSE in this folder

output = struct;
[f0, loc] = pitch(xa, fs,"Method","CEP","Range",[70 450]);
output.fo = f0;
output.tt = loc/fs-0.028;
output.titleStr = "CEP ";
output.filePrefix = "pCEP";
end
\end{verbatim}
}
\caption{Interface function for a cepstrum-based pitch extractor implemented as a MATLAB builtin function.\label{fig:interface}
The constant -0.028 is for time alignment (Unit: second).}
\end{figure}
Figure~\ref{fig:interface} shows an interface function of a MATLAB function to the measuring program.
The pitch extractor is a MATLAB builtin function \textsf{pitch} with option setting \textsf{CEP} to use the cepstrum-based algorithm \cite{noll1967jasa}.
Editing relevant lines in this function provides the interface for any pitch extractors implemented using MATLAB.

\begin{figure}
{\scriptsize
\begin{verbatim}
 audiowrite("testSignal.wav", xa/max(abs(xa))*0.8, fs, ...
     "BitsPerSample",24);
!PATH=$PATH:~/Downloads/opensmile-3.0.1-macos-x64/bin/
!SMILExtract -C prosodyShs.conf -I testSignal.wav -O testSignal.htk
\end{verbatim}
}
\caption{Interface function for an external function in openSMILE.\label{fig:interfaceShell}
We added a function to read the output file which uses HTK \cite{HTK} format.}
\end{figure}
Figure~\ref{fig:interfaceShell} shows an excerpt of the interface function for calling an external pitch extractor.
The pitch extractor is a prosody feature extractor in openSMILE configured using \textsf{prosodyShs.conf} which uses the subharmonic sampling algorithm \cite{hermes1988measurement}.
This excerpt uses the shell escape syntax of MATLAB on macOS.
Editing these lines provide means to evaluate any external pitch extractors.


\bibliography{mesPitch.bib}

\begin{thebibliography}{63}
\def\enquote#1{``#1,''}
\def\plainquote#1{``#1''}
\expandafter\ifx\csname natexlab\endcsname\relax\def\natexlab#1{#1}\fi
\providecommand{\dourl}[1]{\href{http://#1}{\nolinkurl{#1}}}
\providecommand{\bibinfo}[2]{#2}
\providecommand{\noopsort}[1]{}
\providecommand{\switchargs}[2]{#2#1}
  \def\eatspace #1{#1}

\bibitem[{Agiomyrgiannakis \emph{et~al.}(2017)Agiomyrgiannakis, Kawahara, and
  Zen}]{yangVocoder}
\bibinfo{author}{Agiomyrgiannakis, Y.}, \bibinfo{author}{Kawahara, H.},  and
  \bibinfo{author}{Zen, H.} (\textbf{\bibinfo{year}{2017}}).
  \plainquote{\bibinfo{title}{{YANG VOCODER: Yet-ANother-Generalized VOCODER}}}
  \dourl{https://github.com/google/yang_vocoder}, \bibinfo{note}{(Last viewed
  March 17, 2022)}.

\bibitem[{Atal(1972)}]{atal1972automatic}
\bibinfo{author}{Atal, B.~S.} (\textbf{\bibinfo{year}{1972}}).
  \enquote{\bibinfo{title}{Automatic speaker recognition based on pitch
  contours}} \bibinfo{journal}{J. Acoust. Soc. Am.} \textbf{52}(6B),
  \bibinfo{pages}{1687--1697}.

\bibitem[{Behroozmand \emph{et~al.}(2020)Behroozmand, Johari, Bridwell, Hayden,
  Fahey, and Den~Ouden}]{behroozmand2020modulation}
\bibinfo{author}{Behroozmand, R.}, \bibinfo{author}{Johari, K.},
  \bibinfo{author}{Bridwell, K.}, \bibinfo{author}{Hayden, C.},
  \bibinfo{author}{Fahey, D.},  and \bibinfo{author}{Den~Ouden, D.-B.}
  (\textbf{\bibinfo{year}{2020}}). \enquote{\bibinfo{title}{Modulation of vocal
  pitch control through high-definition transcranial direct current stimulation
  of the left ventral motor cortex}} \bibinfo{journal}{Exp. Brain Res.}
  \textbf{238}, \bibinfo{pages}{1525--1535}.

\bibitem[{Behroozmand \emph{et~al.}(2012)Behroozmand, Korzyukov, and
  Larson}]{BEHROOZMAND201289}
\bibinfo{author}{Behroozmand, R.}, \bibinfo{author}{Korzyukov, O.},  and
  \bibinfo{author}{Larson, C.~R.} (\textbf{\bibinfo{year}{2012}}).
  \enquote{\bibinfo{title}{{ERP} correlates of pitch error detection in complex
  tone and voice auditory feedback with missing fundamental}}
  \bibinfo{journal}{Brain Res.} \textbf{1448}, \bibinfo{pages}{89--100},
  \dodoi{https://doi.org/10.1016/j.brainres.2012.02.012}.

\bibitem[{Boersma(1993)}]{boersma1993accurate}
\bibinfo{author}{Boersma, P.} (\textbf{\bibinfo{year}{1993}}).
  \enquote{\bibinfo{title}{Accurate short-term analysis of the fundamental
  frequency and the harmonics-to-noise ratio of a sampled sound}} in
  \emph{\bibinfo{booktitle}{Proc. ICPhS}}, Vol. 17, pp.
  \bibinfo{pages}{97--110}.

\bibitem[{Boersma and Weenink(1992--2022)}]{boersma2011praat}
\bibinfo{author}{Boersma, P.},  and \bibinfo{author}{Weenink, D.}
  (\textbf{\bibinfo{year}{1992--2022}}). \plainquote{\bibinfo{title}{Praat:
  doing phonetics by computer [computer program] version 6.2.10}}
  \dourl{{http://www.praat.org/}}, \bibinfo{note}{(Last viewed March 17,
  2022)}.

\bibitem[{Brooks()}]{brooksVoicevox}
\bibinfo{author}{Brooks, M.} \plainquote{\bibinfo{title}{{VOICEBOX: Speech}
  processing toolbox for {MATLAB}}}
  \dourl{http://www.ee.ic.ac.uk/hp/staff/dmb/voicebox/voicebox.html},
  \bibinfo{note}{(Last viewed March 30, 2022)}.

\bibitem[{Burnett \emph{et~al.}(1997)Burnett, Senner, and
  Larson}]{burnett1997voice}
\bibinfo{author}{Burnett, T.~A.}, \bibinfo{author}{Senner, J.~E.},  and
  \bibinfo{author}{Larson, C.~R.} (\textbf{\bibinfo{year}{1997}}).
  \enquote{\bibinfo{title}{Voice {F0} responses to pitch-shifted auditory
  feedback: a preliminary study}} \bibinfo{journal}{J. Voice} \textbf{11}(2),
  \bibinfo{pages}{202--211}.

\bibitem[{Burrascano \emph{et~al.}(2019)Burrascano, Laureti, Ricci, Terenzi,
  Cecchi, Spinsante, and Piazza}]{burrascano2019swept}
\bibinfo{author}{Burrascano, P.}, \bibinfo{author}{Laureti, S.},
  \bibinfo{author}{Ricci, M.}, \bibinfo{author}{Terenzi, A.},
  \bibinfo{author}{Cecchi, S.}, \bibinfo{author}{Spinsante, S.},  and
  \bibinfo{author}{Piazza, F.} (\textbf{\bibinfo{year}{2019}}).
  \enquote{\bibinfo{title}{A swept-sine pulse compression procedure for an
  effective measurement of intermodulation distortion}} \bibinfo{journal}{IEEE
  Trans. Instrum. Meas.} \textbf{69}(4), \bibinfo{pages}{1708--1719}.

\bibitem[{Camacho(2007)}]{camacho2007swipe}
\bibinfo{author}{Camacho, A.} (\textbf{\bibinfo{year}{2007}}).
  \emph{\bibinfo{title}{SWIPE: A sawtooth waveform inspired pitch estimator for
  speech and music}} (\bibinfo{publisher}{University of Florida Gainesville}).

\bibitem[{Camacho and Harris(2008)}]{camacho2008jasa}
\bibinfo{author}{Camacho, A.},  and \bibinfo{author}{Harris, J.~G.}
  (\textbf{\bibinfo{year}{2008}}). \enquote{\bibinfo{title}{A sawtooth waveform
  inspired pitch estimator for speech and music}} \bibinfo{journal}{J. Acoust.
  Soc. Am.} \textbf{124}(3), \bibinfo{pages}{1638--1652},
  \dodoi{10.1121/1.2951592}.

\bibitem[{Cambridge(1989--2016)}]{HTK}
\bibinfo{author}{Cambridge} (\textbf{\bibinfo{year}{1989--2016}}).
  \plainquote{\bibinfo{title}{{The Hidden Markov Toolkit (HTK)}}}
  \dourl{https://htk.eng.cam.ac.uk/}, \bibinfo{note}{(Last viewed March 31,
  2022)}.

\bibitem[{Chang \emph{et~al.}(2013)Chang, Niziolek, Knight, Nagarajan, and
  Houde}]{houde2013PNAS}
\bibinfo{author}{Chang, E.~F.}, \bibinfo{author}{Niziolek, C.~A.},
  \bibinfo{author}{Knight, R.~T.}, \bibinfo{author}{Nagarajan, S.~S.},  and
  \bibinfo{author}{Houde, J.~F.} (\textbf{\bibinfo{year}{2013}}).
  \enquote{\bibinfo{title}{Human cortical sensorimotor network underlying
  feedback control of vocal pitch}} \bibinfo{journal}{Proc. Natl. Acad. Sci.
  U.S.A.} \textbf{110}(7), \bibinfo{pages}{2653--2658}.

\bibitem[{Das \emph{et~al.}(2020)Das, Smith~III, and Chafe}]{das2020improved}
\bibinfo{author}{Das, O.}, \bibinfo{author}{Smith~III, J.~O.},  and
  \bibinfo{author}{Chafe, C.} (\textbf{\bibinfo{year}{2020}}).
  \enquote{\bibinfo{title}{Improved real-time monophonic pitch tracking with
  the extended complex {Kalman} filter}} \bibinfo{journal}{J. Audio Engineering
  Society} \textbf{68}(1/2), \bibinfo{pages}{78--86}.

\bibitem[{de~Cheveign{\'e} and Kawahara(2002)}]{de2002yin}
\bibinfo{author}{de~Cheveign{\'e}, A.},  and \bibinfo{author}{Kawahara, H.}
  (\textbf{\bibinfo{year}{2002}}). \enquote{\bibinfo{title}{{YIN}, a
  fundamental frequency estimator for speech and music}} \bibinfo{journal}{J.
  Acoust. Soc. Am.} \textbf{111}(4), \bibinfo{pages}{1917--1930}.

\bibitem[{Degottex \emph{et~al.}()Degottex, Kane, Drugman, Raitio, and
  Scherer}]{covarep}
\bibinfo{author}{Degottex, G.}, \bibinfo{author}{Kane, J.},
  \bibinfo{author}{Drugman, T.}, \bibinfo{author}{Raitio, T.},  and
  \bibinfo{author}{Scherer, S.} \plainquote{\bibinfo{title}{Covarep: A
  cooperative voice analysis repository for speech technologies version (after
  1.4.1)}} \dourl{https://github.com/covarep/covarep}, \bibinfo{note}{(Last
  viewed March 30, 2022)}.

\bibitem[{Degottex \emph{et~al.}(2014)Degottex, Kane, Drugman, Raitio, and
  Scherer}]{Degottex2014icassp}
\bibinfo{author}{Degottex, G.}, \bibinfo{author}{Kane, J.},
  \bibinfo{author}{Drugman, T.}, \bibinfo{author}{Raitio, T.},  and
  \bibinfo{author}{Scherer, S.} (\textbf{\bibinfo{year}{2014}}).
  \enquote{\bibinfo{title}{{COVAREP} -- {A} collaborative voice analysis
  repository for speech technologies}} in \emph{\bibinfo{booktitle}{Proc.
  ICASSP}}, pp. \bibinfo{pages}{960--964}, \dodoi{10.1109/ICASSP.2014.6853739}.

\bibitem[{Drugman and Alwan(2011)}]{drugman11interspeech}
\bibinfo{author}{Drugman, T.},  and \bibinfo{author}{Alwan, A.}
  (\textbf{\bibinfo{year}{2011}}). \enquote{\bibinfo{title}{{Joint robust
  voicing detection and pitch estimation based on residual harmonics}}} in
  \emph{\bibinfo{booktitle}{Proc. Interspeech}}, pp.
  \bibinfo{pages}{1973--1976}, \dodoi{10.21437/Interspeech.2011-519}.

\bibitem[{Eyben \emph{et~al.}(2010{\natexlab{a}})Eyben, W{\"o}llmer, and
  Schuller}]{eyben2010opensmile}
\bibinfo{author}{Eyben, F.}, \bibinfo{author}{W{\"o}llmer, M.},  and
  \bibinfo{author}{Schuller, B.} (\textbf{\bibinfo{year}{2010}}{\natexlab{a}}).
  \enquote{\bibinfo{title}{{openSMILE} -- the munich versatile and fast
  open-source audio feature extractor}} in \emph{\bibinfo{booktitle}{Proc. 18th
  ACM Multimedia}}, pp. \bibinfo{pages}{1459--1462}.

\bibitem[{Eyben \emph{et~al.}(2010{\natexlab{b}})Eyben, W{\"o}llmer, and
  Schuller}]{openSmile3}
\bibinfo{author}{Eyben, F.}, \bibinfo{author}{W{\"o}llmer, M.},  and
  \bibinfo{author}{Schuller, B.} (\textbf{\bibinfo{year}{2010}}{\natexlab{b}}).
  \plainquote{\bibinfo{title}{{openSMILE 3.0}: Open-source audio feature
  extraction}} \dourl{https://github.com/audeering/opensmile},
  \bibinfo{note}{(Last viewed March 17, 2022)}.

\bibitem[{Eyben \emph{et~al.}(2010{\natexlab{c}})Eyben, W{\"o}llmer, and
  Schuller}]{eyben2010opensmile2}
\bibinfo{author}{Eyben, F.}, \bibinfo{author}{W{\"o}llmer, M.},  and
  \bibinfo{author}{Schuller, B.} (\textbf{\bibinfo{year}{2010}}{\natexlab{c}}).
  \enquote{\bibinfo{title}{Opensmile: the munich versatile and fast open-source
  audio feature extractor}} in \emph{\bibinfo{booktitle}{Proceedings of the
  18th ACM international conference on Multimedia}}, pp.
  \bibinfo{pages}{1459--1462}.

\bibitem[{Farina(2000)}]{farina2000simultaneous}
\bibinfo{author}{Farina, A.} (\textbf{\bibinfo{year}{2000}}).
  \enquote{\bibinfo{title}{Simultaneous measurement of impulse response and
  distortion with a swept-sine technique}} in \emph{\bibinfo{booktitle}{Audio
  Eng. Soc. Conv. 108}}, \bibinfo{organization}{AES}.

\bibitem[{Gonzalez and Brookes(2014)}]{Gonzalez2014PEFAC}
\bibinfo{author}{Gonzalez, S.},  and \bibinfo{author}{Brookes, M.}
  (\textbf{\bibinfo{year}{2014}}). \enquote{\bibinfo{title}{Pefac - a pitch
  estimation algorithm robust to high levels of noise}}
  \bibinfo{journal}{IEEE/ACM Trans. ASLP} \textbf{22}(2),
  \bibinfo{pages}{518--530}, \dodoi{10.1109/TASLP.2013.2295918}.

\bibitem[{Hain \emph{et~al.}(2000)Hain, Burnett, Kiran, Larson, Singh, and
  Kenney}]{hain2000instructing}
\bibinfo{author}{Hain, T.~C.}, \bibinfo{author}{Burnett, T.~A.},
  \bibinfo{author}{Kiran, S.}, \bibinfo{author}{Larson, C.~R.},
  \bibinfo{author}{Singh, S.},  and \bibinfo{author}{Kenney, M.~K.}
  (\textbf{\bibinfo{year}{2000}}). \enquote{\bibinfo{title}{Instructing
  subjects to make a voluntary response reveals the presence of two components
  to the audio-vocal reflex}} \bibinfo{journal}{Exp. Brain Res.}
  \textbf{130}(2), \bibinfo{pages}{133--141}.

\bibitem[{Hermes(1988)}]{hermes1988measurement}
\bibinfo{author}{Hermes, D.~J.} (\textbf{\bibinfo{year}{1988}}).
  \enquote{\bibinfo{title}{Measurement of pitch by subharmonic summation}}
  \bibinfo{journal}{J. Acoust. Soc. Am.} \textbf{83}(1),
  \bibinfo{pages}{257--264}.

\bibitem[{Kadiri \emph{et~al.}(2021)Kadiri, Alku, and
  Yegnanarayana}]{kadiri2021extraction}
\bibinfo{author}{Kadiri, S.~R.}, \bibinfo{author}{Alku, P.},  and
  \bibinfo{author}{Yegnanarayana, B.} (\textbf{\bibinfo{year}{2021}}).
  \enquote{\bibinfo{title}{Extraction and utilization of excitation information
  of speech: A review}} \bibinfo{journal}{Proceedings of the IEEE} .

\bibitem[{Kawahara(1994)}]{kawahara1994interactions}
\bibinfo{author}{Kawahara, H.} (\textbf{\bibinfo{year}{1994}}).
  \enquote{\bibinfo{title}{Interactions between speech production and
  perception under auditory feedback perturbations on fundamental frequencies}}
  \bibinfo{journal}{J. Acoust. Soc. Jpn. (E)} \textbf{15}(3),
  \bibinfo{pages}{201--202}.

\bibitem[{Kawahara(2016)}]{Kawahara2016}
\bibinfo{author}{Kawahara, H.} (\textbf{\bibinfo{year}{2016}}).
  \enquote{\bibinfo{title}{Aliasing-free {L-F} model and its application to an
  interactive {MATLAB} tool and test signal generation for speech analysis
  procedures}} in \emph{\bibinfo{booktitle}{9th ISCA Speech Synthesis
  Workshop}}, pp. \bibinfo{pages}{123--123}.

\bibitem[{Kawahara(2017{\natexlab{a}})}]{HidekiKawahara2017}
\bibinfo{author}{Kawahara, H.} (\textbf{\bibinfo{year}{2017}}{\natexlab{a}}).
  \enquote{\bibinfo{title}{Application of time-frequency representations of
  aperiodicity and instantaneous frequency for detailed analysis of filled
  pauses}} \bibinfo{journal}{Journal of the Phonetic Society of Japan}
  \textbf{21}(3), \bibinfo{pages}{63--73},
  \dodoi{10.24467/onseikenkyu.21.3_63}.

\bibitem[{Kawahara(2017{\natexlab{b}})}]{legacySTRAIGHT}
\bibinfo{author}{Kawahara, H.} (\textbf{\bibinfo{year}{2017}}{\natexlab{b}}).
  \plainquote{\bibinfo{title}{{legacy-STRAIGHT}}}
  \dourl{https://github.com/HidekiKawahara/legacy_STRAIGHT},
  \bibinfo{note}{(Last viewed March 17, 2022)}.

\bibitem[{Kawahara(2017{\natexlab{c}})}]{ninjalGit}
\bibinfo{author}{Kawahara, H.} (\textbf{\bibinfo{year}{2017}}{\natexlab{c}}).
  \plainquote{\bibinfo{title}{{YANGstraight source}}}
  \dourl{https://github.com/HidekiKawahara/YANGstraight_source},
  \bibinfo{note}{(Last viewed March 17, 2022)}.

\bibitem[{Kawahara \emph{et~al.}(2005)Kawahara, de~Cheveigné, Banno,
  Takahashi, and Irino}]{kawahara05_interspeech}
\bibinfo{author}{Kawahara, H.}, \bibinfo{author}{de~Cheveigné, A.},
  \bibinfo{author}{Banno, H.}, \bibinfo{author}{Takahashi, T.},  and
  \bibinfo{author}{Irino, T.} (\textbf{\bibinfo{year}{2005}}).
  \enquote{\bibinfo{title}{{Nearly defect-free F0 trajectory extraction for
  expressive speech modifications based on STRAIGHT}}} in
  \emph{\bibinfo{booktitle}{Proc. Interspeech}}, pp. \bibinfo{pages}{537--540},
  \dodoi{10.21437/Interspeech.2005-335}.

\bibitem[{{Kawahara} \emph{et~al.}(1996){Kawahara}, {Kato}, and
  {Williams}}]{kawahara1996icslp}
\bibinfo{author}{{Kawahara}, H.}, \bibinfo{author}{{Kato}, H.},  and
  \bibinfo{author}{{Williams}, J.~C.} (\textbf{\bibinfo{year}{1996}}).
  \enquote{\bibinfo{title}{Effects of auditory feedback on {F0} trajectory
  generation}} in \emph{\bibinfo{booktitle}{Proc. ICSLP}}, Vol. 1, pp.
  \bibinfo{pages}{287--290}, \dodoi{10.1109/ICSLP.1996.607105}.

\bibitem[{Kawahara \emph{et~al.}(1999)Kawahara, Masuda-Katsuse, and
  de~Cheveign{\'e}}]{kawahara1999spcom}
\bibinfo{author}{Kawahara, H.}, \bibinfo{author}{Masuda-Katsuse, I.},  and
  \bibinfo{author}{de~Cheveign{\'e}, A.} (\textbf{\bibinfo{year}{1999}}).
  \enquote{\bibinfo{title}{{Restructuring speech representations using a
  pitch-adaptive time-frequency smoothing and an instantaneous-frequency-based
  F0 extraction}}} \bibinfo{journal}{Speech Communication} \textbf{27}(3-4),
  \bibinfo{pages}{187--207}.

\bibitem[{Kawahara \emph{et~al.}(2021{\natexlab{a}})Kawahara, Matsui, Yatabe,
  Sakakibara, Tsuzaki, Morise, and Irino}]{kawahara2021APSIPA}
\bibinfo{author}{Kawahara, H.}, \bibinfo{author}{Matsui, T.},
  \bibinfo{author}{Yatabe, K.}, \bibinfo{author}{Sakakibara, K.-I.},
  \bibinfo{author}{Tsuzaki, M.}, \bibinfo{author}{Morise, M.},  and
  \bibinfo{author}{Irino, T.} (\textbf{\bibinfo{year}{2021}}{\natexlab{a}}).
  \enquote{\bibinfo{title}{Implementation of interactive tools for
  investigating fundamental frequency response of voiced sounds to auditory
  stimulation}} in \emph{\bibinfo{booktitle}{Proc. APSIPA ASC}}, pp.
  \bibinfo{pages}{897--903}.

\bibitem[{Kawahara \emph{et~al.}(2021{\natexlab{b}})Kawahara, Matsui, Yatabe,
  Sakakibara, Tsuzaki, Morise, and Irino}]{kawahara2021interspeech}
\bibinfo{author}{Kawahara, H.}, \bibinfo{author}{Matsui, T.},
  \bibinfo{author}{Yatabe, K.}, \bibinfo{author}{Sakakibara, K.-I.},
  \bibinfo{author}{Tsuzaki, M.}, \bibinfo{author}{Morise, M.},  and
  \bibinfo{author}{Irino, T.} (\textbf{\bibinfo{year}{2021}}{\natexlab{b}}).
  \enquote{\bibinfo{title}{Mixture of orthogonal sequences made from extended
  time-stretched pulses enables measurement of involuntary voice fundamental
  frequency response to pitch perturbation}} in \emph{\bibinfo{booktitle}{Proc.
  Interspeech 2021}}, pp. \bibinfo{pages}{3206--3210}.

\bibitem[{Kawahara \emph{et~al.}(2008{\natexlab{a}})Kawahara, Morise,
  Takahashi, Nisimura, Banno, and Irino}]{kawahara08_spkd}
\bibinfo{author}{Kawahara, H.}, \bibinfo{author}{Morise, M.},
  \bibinfo{author}{Takahashi, T.}, \bibinfo{author}{Nisimura, R.},
  \bibinfo{author}{Banno, H.},  and \bibinfo{author}{Irino, T.}
  (\textbf{\bibinfo{year}{2008}}{\natexlab{a}}). \enquote{\bibinfo{title}{{A
  unified approach for F0 extraction and aperiodicity estimation based on a
  temporally stable power spectral representation}}} in
  \emph{\bibinfo{booktitle}{Proc. ISCA ITRW on Speech Analysis and Processing
  for Knowledge Discovery}}, p. \bibinfo{pages}{paper 043}.

\bibitem[{Kawahara \emph{et~al.}(2008{\natexlab{b}})Kawahara, Morise,
  Takahashi, Nisimura, Irino, and Banno}]{kawahara2008icassp}
\bibinfo{author}{Kawahara, H.}, \bibinfo{author}{Morise, M.},
  \bibinfo{author}{Takahashi, T.}, \bibinfo{author}{Nisimura, R.},
  \bibinfo{author}{Irino, T.},  and \bibinfo{author}{Banno, H.}
  (\textbf{\bibinfo{year}{2008}}{\natexlab{b}}).
  \enquote{\bibinfo{title}{Tandem-straight: A temporally stable power spectral
  representation for periodic signals and applications to interference-free
  spectrum, f0, and aperiodicity estimation}} in
  \emph{\bibinfo{booktitle}{Proc. ICASSP}}, pp. \bibinfo{pages}{3933--3936},
  \dodoi{10.1109/ICASSP.2008.4518514}.

\bibitem[{Kawahara \emph{et~al.}(2017)Kawahara, Sakakibara, Morise, Banno, and
  Toda}]{kawahara2017accurate}
\bibinfo{author}{Kawahara, H.}, \bibinfo{author}{Sakakibara, K.-I.},
  \bibinfo{author}{Morise, M.}, \bibinfo{author}{Banno, H.},  and
  \bibinfo{author}{Toda, T.} (\textbf{\bibinfo{year}{2017}}).
  \enquote{\bibinfo{title}{Accurate estimation of fo and aperiodicity based on
  periodicity detector residuals and deviations of phase derivatives}} in
  \emph{\bibinfo{booktitle}{Proc. APSIPA ASC}}, pp. \bibinfo{pages}{12--15}.

\bibitem[{Kawahara and Yatabe(2021)}]{kawahara2021icassp}
\bibinfo{author}{Kawahara, H.},  and \bibinfo{author}{Yatabe, K.}
  (\textbf{\bibinfo{year}{2021}}). \enquote{\bibinfo{title}{Cascaded all-pass
  filters with randomized center frequencies and phase polarity for acoustic
  and speech measurement and data augmentation}} \bibinfo{journal}{Proc.
  ICASSP2021} \bibinfo{pages}{306--310}.

\bibitem[{Kawahara \emph{et~al.}(2022)Kawahara, Yatabe, Sakakibara, Kitamura,
  Banno, and Morise}]{kawahara2022IS}
\bibinfo{author}{Kawahara, H.}, \bibinfo{author}{Yatabe, K.},
  \bibinfo{author}{Sakakibara, K.-I.}, \bibinfo{author}{Kitamura, T.},
  \bibinfo{author}{Banno, H.},  and \bibinfo{author}{Morise, M.}
  (\textbf{\bibinfo{year}{2022}}). \enquote{\bibinfo{title}{An objective test
  tool for pitch extractors' response attributes}} in
  \emph{\bibinfo{booktitle}{Proc. Interspeech}}, \bibinfo{note}{(submitted)}.

\bibitem[{Kim \emph{et~al.}(2018)Kim, Salamon, Li, and Bello}]{kim2018crepe}
\bibinfo{author}{Kim, J.~W.}, \bibinfo{author}{Salamon, J.},
  \bibinfo{author}{Li, P.},  and \bibinfo{author}{Bello, J.~P.}
  (\textbf{\bibinfo{year}{2018}}). \enquote{\bibinfo{title}{{CREPE: A}
  convolutional representation for pitch estimation}} in
  \emph{\bibinfo{booktitle}{Proc. ICASSP}}, pp. \bibinfo{pages}{161--165}.

\bibitem[{Kleijn(2003)}]{kleijn2003signal}
\bibinfo{author}{Kleijn, W.~B.} (\textbf{\bibinfo{year}{2003}}).
  \enquote{\bibinfo{title}{Signal processing representations of speech}}
  \bibinfo{journal}{IEICE TRANSACTIONS on Information and Systems}
  \textbf{86}(3), \bibinfo{pages}{359--376}.

\bibitem[{Larson and Robin(2016)}]{larson2016sensory}
\bibinfo{author}{Larson, C.~R.},  and \bibinfo{author}{Robin, D.~A.}
  (\textbf{\bibinfo{year}{2016}}). \enquote{\bibinfo{title}{Sensory processing:
  {Advances} in understanding structure and function of pitch-shifted auditory
  feedback in voice control}} \bibinfo{journal}{AIMS Neurosci.} \textbf{3}(1),
  \bibinfo{pages}{22--39}, \dodoi{10.3934/Neuroscience.2016.1.22}.

\bibitem[{Maekawa(2003)}]{maekawa03_sspr}
\bibinfo{author}{Maekawa, K.} (\textbf{\bibinfo{year}{2003}}).
  \enquote{\bibinfo{title}{{Corpus of spontaneous Japanese: its design and
  evaluation}}} in \emph{\bibinfo{booktitle}{Proc. ISCA/IEEE Workshop on
  Spontaneous Speech Processing and Recognition}}, p. \bibinfo{pages}{paper
  MMO2}.

\bibitem[{McAulay and Quatieri(1986)}]{mcaulay1986speech}
\bibinfo{author}{McAulay, R.},  and \bibinfo{author}{Quatieri, T.}
  (\textbf{\bibinfo{year}{1986}}). \enquote{\bibinfo{title}{Speech
  analysis/synthesis based on a sinusoidal representation}}
  \bibinfo{journal}{IEEE Transactions on Acoustics, Speech, and Signal
  Processing} \textbf{34}(4), \bibinfo{pages}{744--754}.

\bibitem[{Morise(2017)}]{morise17b_interspeech}
\bibinfo{author}{Morise, M.} (\textbf{\bibinfo{year}{2017}}).
  \enquote{\bibinfo{title}{{Harvest: A} high-performance fundamental frequency
  estimator from speech signals}} in \emph{\bibinfo{booktitle}{Proc.
  Interspeech}}, pp. \bibinfo{pages}{2321--2325},
  \dodoi{10.21437/Interspeech.2017-68}.

\bibitem[{Morise \emph{et~al.}(2016)Morise, Yokomori, and
  Ozawa}]{morise2016world}
\bibinfo{author}{Morise, M.}, \bibinfo{author}{Yokomori, F.},  and
  \bibinfo{author}{Ozawa, K.} (\textbf{\bibinfo{year}{2016}}).
  \enquote{\bibinfo{title}{{WORLD:} {A} vocoder-based high-quality speech
  synthesis system for real-time applications}} \bibinfo{journal}{IEICE Trans.
  Information and Systems} \textbf{99}(7), \bibinfo{pages}{1877--1884}.

\bibitem[{Noll(1967)}]{noll1967jasa}
\bibinfo{author}{Noll, A.~M.} (\textbf{\bibinfo{year}{1967}}).
  \enquote{\bibinfo{title}{Cepstrum pitch determination}} \bibinfo{journal}{J.
  Acoust. Soc. Am.} \textbf{41}(2), \bibinfo{pages}{293--309},
  \dodoi{10.1121/1.1910339}.

\bibitem[{Patel \emph{et~al.}(2016)Patel, Lodhavia, Frankford, Korzyukov, and
  Larson}]{PATEL2016772.e33}
\bibinfo{author}{Patel, S.}, \bibinfo{author}{Lodhavia, A.},
  \bibinfo{author}{Frankford, S.}, \bibinfo{author}{Korzyukov, O.},  and
  \bibinfo{author}{Larson, C.~R.} (\textbf{\bibinfo{year}{2016}}).
  \enquote{\bibinfo{title}{Vocal and neural responses to unexpected changes in
  voice pitch auditory feedback during register transitions}}
  \bibinfo{journal}{J. Voice} \textbf{30}(6),
  \bibinfo{pages}{772.e33--772.e40},
  \dodoi{https://doi.org/10.1016/j.jvoice.2015.11.012}.

\bibitem[{Patterson(1987)}]{patterson1987pulse}
\bibinfo{author}{Patterson, R.~D.} (\textbf{\bibinfo{year}{1987}}).
  \enquote{\bibinfo{title}{A pulse ribbon model of monaural phase perception}}
  \bibinfo{journal}{The Journal of the Acoustical Society of America}
  \textbf{82}(5), \bibinfo{pages}{1560--1586}.

\bibitem[{Peng \emph{et~al.}(2021)Peng, Lin, Chang, Jones, Jia, Chen, Liu, and
  Liu}]{peng2021causal}
\bibinfo{author}{Peng, D.}, \bibinfo{author}{Lin, Q.}, \bibinfo{author}{Chang,
  Y.}, \bibinfo{author}{Jones, J.~A.}, \bibinfo{author}{Jia, G.},
  \bibinfo{author}{Chen, X.}, \bibinfo{author}{Liu, P.},  and
  \bibinfo{author}{Liu, H.} (\textbf{\bibinfo{year}{2021}}).
  \enquote{\bibinfo{title}{A causal role of the cerebellum in auditory feedback
  control of vocal production}} \bibinfo{journal}{Cerebellum}
  \bibinfo{pages}{1--12}.

\bibitem[{Schroeder(1970)}]{schroeder1970synthesis}
\bibinfo{author}{Schroeder, M.} (\textbf{\bibinfo{year}{1970}}).
  \enquote{\bibinfo{title}{Synthesis of low-peak-factor signals and binary
  sequences with low autocorrelation}} \bibinfo{journal}{IEEE Trans. Inf.
  Theory} \textbf{16}(1), \bibinfo{pages}{85--89}.

\bibitem[{Schuller and Batliner(2013)}]{schuller2013computational}
\bibinfo{author}{Schuller, B.},  and \bibinfo{author}{Batliner, A.}
  (\textbf{\bibinfo{year}{2013}}). \emph{\bibinfo{title}{Computational
  paralinguistics: emotion, affect and personality in speech and language
  processing}} (\bibinfo{publisher}{John Wiley \& Sons}).

\bibitem[{Sivasankar \emph{et~al.}(2005)Sivasankar, Bauer, Babu, and
  Larson}]{larson2005Jasa}
\bibinfo{author}{Sivasankar, M.}, \bibinfo{author}{Bauer, J.~J.},
  \bibinfo{author}{Babu, T.},  and \bibinfo{author}{Larson, C.~R.}
  (\textbf{\bibinfo{year}{2005}}). \enquote{\bibinfo{title}{Voice responses to
  changes in pitch of voice or tone auditory feedback}} \bibinfo{journal}{J.
  Acoust. Soc. Am.} \textbf{117}(2), \bibinfo{pages}{850--857},
  \dodoi{10.1121/1.1849933}.

\bibitem[{Stan \emph{et~al.}(2002)Stan, Embrechts, and
  Archambeau}]{stan2002comparison}
\bibinfo{author}{Stan, G.-B.}, \bibinfo{author}{Embrechts, J.-J.},  and
  \bibinfo{author}{Archambeau, D.} (\textbf{\bibinfo{year}{2002}}).
  \enquote{\bibinfo{title}{Comparison of different impulse response measurement
  techniques}} \bibinfo{journal}{J. Audio Eng. Soc.} \textbf{50}(4),
  \bibinfo{pages}{249--262}.

\bibitem[{Talkin()}]{googleREAPERgit}
\bibinfo{author}{Talkin, D.} \plainquote{\bibinfo{title}{{REAPER: Robust} epoch
  and pitch estimator}} \dourl{https://github.com/google/REAPER},
  \bibinfo{note}{(Last viewed March 21, 2022)}.

\bibitem[{Talkin(1995)}]{talkin1995robust}
\bibinfo{author}{Talkin, D.} (\textbf{\bibinfo{year}{1995}}).
  \enquote{\bibinfo{title}{A robust algorithm for pitch tracking {(RAPT)}}} in
  \emph{\bibinfo{booktitle}{Speech coding and synthesis}}, edited by
  \bibinfo{editor}{D.~Talkin} and \bibinfo{editor}{W.~B. Kleijn}
  (\bibinfo{publisher}{Elsevier Science}), pp. \bibinfo{pages}{497--512}.

\bibitem[{The~Mathworks(2022)}]{MATLAB:R2022a}
\bibinfo{author}{The~Mathworks, I.} (\textbf{\bibinfo{year}{2022}}).
  \emph{\bibinfo{title}{{MATLAB version 9.12.0.1884302 (R2022a)}}},
  \bibinfo{organization}{The Mathworks, Inc.}, \bibinfo{address}{Natick,
  Massachusetts}.

\bibitem[{Titze(1993)}]{Titze1993}
\bibinfo{author}{Titze, I.~R.} (\textbf{\bibinfo{year}{1993}}).
  \enquote{\bibinfo{title}{Comparison of {Fo} extraction methods for
  high-precision voice perturbation measurements an optimizer-simulator for
  phonosurgery view project}} \bibinfo{journal}{Article in Journal of Speech
  and Hearing Research} \textbf{36}, \bibinfo{pages}{1120--1133},
  \dourl{https://www.researchgate.net/publication/15083731},
  \dodoi{10.1044/jshr.3606.1120}.

\bibitem[{Tourville \emph{et~al.}(2008)Tourville, Reilly, and
  Guenther}]{TOURVILLE20081429}
\bibinfo{author}{Tourville, J.~A.}, \bibinfo{author}{Reilly, K.~J.},  and
  \bibinfo{author}{Guenther, F.~H.} (\textbf{\bibinfo{year}{2008}}).
  \enquote{\bibinfo{title}{Neural mechanisms underlying auditory feedback
  control of speech}} \bibinfo{journal}{NeuroImage} \textbf{39}(3),
  \bibinfo{pages}{1429 -- 1443},
  \dodoi{https://doi.org/10.1016/j.neuroimage.2007.09.054}.

\bibitem[{V\"alim\"aki \emph{et~al.}(2013)V\"alim\"aki, Lehtonen, and
  Takanen}]{valimaki2013ieetr}
\bibinfo{author}{V\"alim\"aki, V.}, \bibinfo{author}{Lehtonen, H.~M.},  and
  \bibinfo{author}{Takanen, M.} (\textbf{\bibinfo{year}{2013}}).
  \enquote{\bibinfo{title}{A perceptual study on velvet noise and its variants
  at different pulse densities}} \bibinfo{journal}{IEEE Trans. Audio, Speech,
  Lang. Process.} \textbf{21}(7), \bibinfo{pages}{1481--1488},
  \dodoi{10.1109/TASL.2013.2255281}.

\bibitem[{Zarate \emph{et~al.}(2010)Zarate, Wood, and
  Zatorre}]{zarate2010neural}
\bibinfo{author}{Zarate, J.~M.}, \bibinfo{author}{Wood, S.},  and
  \bibinfo{author}{Zatorre, R.~J.} (\textbf{\bibinfo{year}{2010}}).
  \enquote{\bibinfo{title}{Neural networks involved in voluntary and
  involuntary vocal pitch regulation in experienced singers}}
  \bibinfo{journal}{Neuropsychologia} \textbf{48}(2),
  \bibinfo{pages}{607--618}.

\end{thebibliography}









\end{document}